\documentclass[10pt,journal,compsoc]{IEEEtran}
\AtEndDocument{\par\leavevmode}

\usepackage{cite}
\usepackage{graphicx} %For images
\usepackage[caption=false]{subfig} %For sub-figures
\usepackage{color} %For TODO colors
\usepackage{flushend} %To balance the two columns
\usepackage{multirow} %Merging cells in tables
\usepackage[export]{adjustbox} %Border around author image

\usepackage{pifont}% http://ctan.org/pkg/pifont , For tick marks and crosses
\newcommand{\cmark}{\ding{51}}%
\newcommand{\xmark}{\ding{53}}%

\usepackage[ruled,norelsize,linesnumbered]{algorithm2e}
\makeatletter
\newcommand{\removelatexerror}{\let\@latex@error\@gobble}
\makeatother

\usepackage{filecontents}

\usepackage{amsbsy}
\usepackage{amsfonts} % For mathbb symbols

\newcommand{\tuplE}[3]{\ensuremath{\langle #1, #2, #3 \rangle}}

\begin{document}
\title{An Adaptive Parallel Algorithm for Computing Connected Components}

\author{
    %\IEEEauthorblockN{Chirag~Jain\IEEEauthorrefmark{1}, Patrick~Flick\IEEEauthorrefmark{1}, Tony~Pan\IEEEauthorrefmark{1}, Oded~Green\IEEEauthorrefmark{1}, Srinivas~Aluru\IEEEauthorrefmark{1}}
    
    \IEEEauthorblockN{Chirag~Jain, Patrick~Flick, Tony~Pan, Oded~Green, Srinivas~Aluru}
    \\{Georgia Institute of Technology, Atlanta, GA 30332, USA
    \\\{cjain, patrick.flick, tony.pan, ogreen\}@gatech.edu, {aluru@cc.gatech.edu}}
}
\markboth{IEEE Transactions on Parallel and Distributed Systems VOL XX}%
{Jain \MakeLowercase{\textit{et al.}}: Parallel Distributed Memory Connectivity Algorithm for Large Undirected Graphs}

\IEEEtitleabstractindextext{%
\begin{abstract}
We present an efficient distributed memory parallel algorithm for computing connected components in undirected graphs based on Shiloach-Vishkin's PRAM approach. We discuss multiple optimization techniques that reduce communication volume as well as load-balance the algorithm. We also note that the efficiency of the parallel graph connectivity algorithm depends on the underlying graph topology. Particularly for short diameter graph components, we observe that parallel Breadth First Search (BFS) method offers better performance. However, running parallel BFS is not efficient for computing large diameter components or large number of small components.  To address this challenge, we employ a heuristic that allows the algorithm to quickly predict the type of the network by computing the degree distribution and follow the optimal hybrid route. Using large graphs with diverse topologies from domains including metagenomics, web crawl, social graph and road networks, we show that our hybrid implementation is efficient and scalable for each of the graph types. Our approach achieves a runtime of 215 seconds using 32K cores of Cray XC30 for a metagenomic graph with over 50 billion edges. When compared against the previous state-of-the-art method, we see performance improvements up to 24x.
\end{abstract}

}
\maketitle
\IEEEdisplaynontitleabstractindextext
\IEEEpeerreviewmaketitle

\section{Introduction}
Computing connected components in undirected graphs is a fundamental problem in graph analytics. The sizes of graph data collections continue to grow in multiple scientific domains, motivating the need for high performance distributed memory parallel graph algorithms, especially for large networks that cannot fit into the memory of a single compute node. For a graph $G(V,E)$ with $n$ vertices and $m$ edges, two vertices belong to the same \emph{connected component} if and only if there is a path between the two vertices in $G$. Sequentially, this problem can be solved in linear $O(m + n)$ time, e.g. by using one of the following two approaches. One approach is to use graph traversal algorithms, i.e., either Breadth First (BFS) or Depth First Search (DFS). A single traversal is necessary for each connected component in the graph. Another technique is to use a union-find based algorithm, where each vertex is initially assumed to be a different graph component and components connected by an edge are iteratively merged.  

Parallel BFS traversal algorithms have been invented that are work-optimal and practical on distributed memory systems for small-world graphs~\cite{bulucc2011parallel,beamer2013distributed}. While parallel BFS algorithms have been optimized for traversing a short diameter big graph component, they can be utilized for finding connected components.  
However, connectivity can be determined for only one component at a time, as BFS cannot merge the multiple partial search trees in the same component that are likely to arise during concurrent runs.  For an undirected graph with a large number of small components, parallel BFS thus has limited utility.  On the other hand, BFS is an efficient technique for scale-free networks that are characterized by having one dominant short diameter component.

The classic Shiloach-Vishkin (SV) algorithm \cite{shiloach1982logn}, a widely known PRAM algorithm for computing connectivity, simultaneously computes connectivity of all the vertices and promises convergence in logarithmic iterations, making it suitable for components with large diameter, as well as for graphs with a large number of small sized components. Note that compared to simple label propagation techniques, the SV algorithm bounds the number of iterations to $O(\log{n})$ instead of $O(n)$, where each iteration requires $O(m + n)$ work. In this work, we provide a novel edge-based parallel algorithm for distributed memory systems based on the SV approach. We also propose optimizations to reduce data volume and balance  load as the iterations progress.

To achieve the best performance for different graph topologies, we introduce a dynamic pre-processing phase to our algorithm that guides the algorithm selection at runtime. In this phase, we try to classify the graph as scale-free by estimating the goodness of fit of its degree distribution to a power-law curve. If and only if the graph is determined to be scale-free, we execute one BFS traversal iteration from a single root to find the largest connected component with high probability, before switching to the SV algorithm to process the remaining graph.  While the pre-processing phase introduces some overhead, we are able to improve the overall performance by using a combination of parallel BFS and SV algorithms, with minimal parameter tuning.

Our primary application driver is metagenomic assembly, where de Bruijn graphs are used for reconstructing, from DNA sequencer outputs, constituent genomes in a metagenome\cite{compeau2011apply}.  A recent scientific study showed that high species-level heterogeneity in metagenomic data sets leads to a large number of weakly connected components, each of which can be processed as independent de Bruijn graphs \cite{howe2014tackling}.  This coarse grained data parallelism motivated our efforts in finding connected components in large metagenomic de Bruijn graphs.  However, our work is applicable to graphs from domains beyond bioinformatics.  

In this study, we cover a diverse set of graphs, both small world and large diameter, to highlight that our algorithm can serve as a general solution to computing connected components for undirected graphs. We experimentally evaluate our algorithm on de Bruijn graphs from publicly available metagenomic samples, road networks of the United States and European Union, scale-free networks from the internet, as well as Kronecker graphs from the Graph500 benchmark \cite{murphy2010introducing}.  The graphs range in edge count from 82 million to 54 billion.   Even though we focus on computing connected components in undirected graphs, ideas discussed in this work are applicable to finding strongly connected components in directed graphs as well.  Our C++ and MPI-based implementation is available as open source at https://github.com/ParBLiSS/parconnect.

To summarize the contributions of this paper:
\begin{itemize}
\item We provide a new scalable strategy to adapt the Shiloach-Vishkin PRAM connectivity algorithm to distributed memory parallel systems. 
\item We discuss and evaluate a novel and efficient dynamic approach to compute weakly connected components on a variety of graphs, with small and large diameters.
\item We demonstrate the scalability of our algorithm by computing the connectivity of the de Bruijn graph for a large metagenomic dataset with 1.8 billion DNA sequences and 54 billion edges in less than 4 minutes using 32K cores.
\item Depending on the underlying graph topology, we see variable performance improvements up to 24x when compared against the state-of-the-art parallel connectivity algorithm.
\end{itemize}

\section{Related Work}
Due to its broad applicability, there have been numerous efforts to parallelize the connected component labeling problem. Hirschberg {\it et al.} \cite{hirschberg1979computing} presented a CREW\footnote{CREW = Concurrent Read Exclusive Write} PRAM algorithm that runs in $O(\log^{2} n)$ time and does $O(n^{2}\log n)$ work, while Shiloach and Vishkin \cite{shiloach1982logn} presented an improved version assuming a CRCW\footnote{CRCW = Concurrent Read Concurrent Write} PRAM that runs in $O(\log n)$ time using $O(m+n)$ processors. 
As our parallel SV algorithm is based on this approach, we summarize the SV algorithm in separate subsection. Krishnamurthy {\it et al.} \cite{Krishnamurthy94connectedcomponents} made the first attempt to adapt SV algorithm to distributed memory machines. 
However, their method is restricted to mesh graphs, which they could naturally partition among the processes \cite{buvs2001parallel}. Goddard {\it et al.} \cite{goddard1994connected} discussed a practical implementation of SV algorithm for distributed machines with mesh network topology. 
Their method, however, was shown to exhibit poor scalability beyond 16 processors for sparse graphs \cite{gregor2005parallel}. 

Bader {\it el al.} \cite{bader2004fast} and Patwary {\it et al.} \cite{patwary2012multi} discussed shared memory multi-threaded parallel implementations to compute spanning forest and connected components on sparse and irregular graphs. Recently, Shun {\it et al.} \cite{shun2014simple} reported a work optimal implementation for the same programming model. Note that these solutions are not applicable for distributed memory environments due to high frequency of remote memory accesses. Cong {\it et al.} \cite{cong2014fast} proposed a parallel technique for solving the connectivity problem on a single GPU.

There have been several recent parallel algorithms for computing the breadth-first search (BFS) traversal on distributed memory systems \cite{bulucc2011parallel,beamer2013distributed,ueno2012highly}. However, parallel BFS does not serve as an efficient, stand-alone method for computing connectivity.  There are also several large-scale distributed graph analytics frameworks that can solve the connectivity problem in large graphs, including GraphX \cite{gonzalez2014graphx}, PowerLyra \cite{chen2015powerlyra}, PowerGraph \cite{gonzalez2012powergraph}, and GraphLab \cite{low2012distributed}. Iverson {\it et al.}~\cite{iverson2015evaluation} proposed a distributed-memory  connectivity algorithm using successive graph contraction operations, however, the strong scalability demonstrated for this method was limited to 32 cores. 

Slota {\it et al.} \cite{slota2014bfs} proposed a shared memory parallel \emph{Multistep} method that combines parallel BFS and label propagation (LP) technique and was reported to perform better than using BFS or LP alone. In their \emph{Multistep} method, BFS is first used to label the largest component before using the LP algorithm to label the remaining components. More recently, they proposed a distributed memory parallel implementation of this method and showed impressive speedups against the existing parallel graph processing frameworks \cite{slota2016high}. However, their algorithm design and experimental datasets are restricted to graphs which contain a single massive connected component.  While our algorithm likewise employs a combination of algorithms, in contrast to \emph{MultiStep}, we use BFS and our novel SV implementation, and determine dynamically at runtime whether the BFS should be executed.

%Recently, it was shown that using the underlying graph structure for algorithm selection at runtime, it is possible to get good resource utilization for betweenness centrality on GPUs \cite{mclaughlin2014scalable}. 
This paper is an extension of our previous work~\cite{flick2015parallel} which described a parallel connected components algorithm based on the SV approach for large diameter metagenomic graphs. Here we propose a hybrid approach using both BFS and SV and present a generalized efficient algorithm for finding connected components in arbitrary undirected graphs. We show that using runtime algorithm selection and our SV implementation, our method generalizes to diverse graph topologies and achieves superior performance.

\subsubsection*{The Shiloach-Vishkin Algorithm}
\begin{figure*}%
\subfloat[]{
\renewcommand{\arraystretch}{1.05}
\begin{tabular}{|p{1.4cm}|p{4.5cm}|p{3.6cm}|} \hline
\bfseries Symbol	&	\bfseries Description & \bfseries Definition \\ \hline\hline

$V$						&	Vertices in graph $G$									&													\\ 
$E$						&	Edges in graph $G$										&													\\ 
$\tuplE{p}{q}{r}$		& 	Tuple 													& $p,q,r \in \mathbb{Z}$							\\  
$\mathcal{A}_{i}$ 		& 	Array of tuples in iteration i 							& 													\\
$\mathcal{P}_{i}$ 		&	Unique partitions										& $\{p \mid \tuplE{p}{q}{r} \in \mathcal{A}_i\}$	\\
$\mathcal{PB}_{i}(p)$	&	Partition bucket for partition $p$						& $\{\tuplE{\hat{p}}{q}{r} \in \mathcal{A}_i \mid \hat{p}=p\}$	\\
$\mathcal{VB}_{i}(u)$	&	Vertex bucket for vertex $u$							& $\{\tuplE{p}{q}{r} \in \mathcal{A}_i \mid r=u\}$	\\
$\mathcal{V}_{i}(p)$	&	Vertex members in partition $p$							& $\{r \mid \tuplE{p}{q}{r} \in \mathcal{PB}_i(p)\}$\\
$\mathcal{C}_{i}(p)$	&	Candidate partitions for partition $p$					& $\{q \mid \tuplE{p}{q}{r} \in \mathcal{PB}_i(p)\}$\\
$\mathcal{M}_{i}(u)$	&	Partitions in vertex bucket for vertex $u$				& $\{p \mid \tuplE{p}{q}{r} \in \mathcal{VB}_i(u)\}$\\
$\mathcal{N}_{i}(p)$	&	Neighborhood partitions of partition $p$				& $\cup_{u \in \mathcal{V}_i(p)}{\mathcal{M}_i(u)}$	\\
\hline
\end{tabular}
\label{tab:symbols}
}
\hfill
\subfloat[]{
\raisebox{-.43\height}{\includegraphics[width=2.3in]{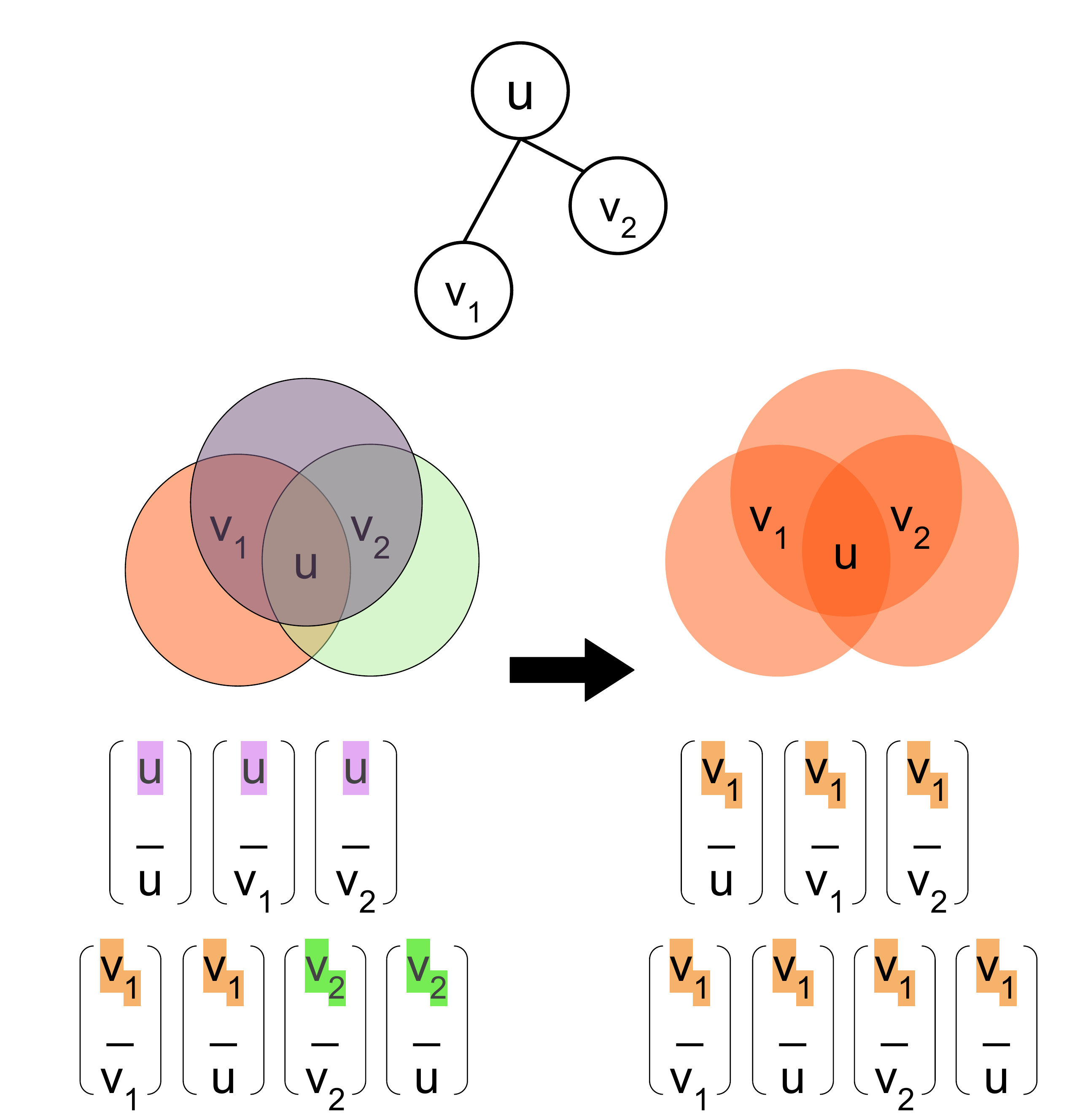}}
\label{fig:algExample}
}
\caption{(a) Summary of the notations used in Section \ref{sec:algo_main}. (b) Initialization of array $\mathcal{A}$ for a small connected component with three vertices $u, v_1, v_2$ in our algorithm. Partitions are highlighted using different shades. Desired solution, assuming $v_1 = \min{(u, v_1, v_2)}$, shown on the right will be to have all three vertices in a single component $v_1$. Accordingly, all the tuples associated with this component should contain the equal partition id $v_1$.}
\end{figure*}
The Shiloach-Vishkin connectivity algorithm was designed assuming a PRAM model. It begins with singleton trees corresponding to each vertex in the graph and maintains this auxiliary structure of rooted directed trees to keep track of the connected components discovered so far during the execution. Within each iteration, there are two phases referred to as $shortcutting$ and $hooking$. $Shortcutting$ involves collapsing the trees using pointer doubling. On the other hand, $hooking$ connects two different connected components when they share an edge in the input graph. This algorithm requires $O(\log{n})$ iterations each taking constant time. Since this approach uses $O(m+n)$ processors, the total work complexity is $O((m+n)\cdot\log n)$.
\section{Algorithm}\label{sec:algo_main}
\subsection{Parallel SV Algorithm}

%----------
% Notations
%----------
\subsubsection{Notations}
\label{sec:algo_defs}
%List of notations being used in the algorithm and the example figure

Given an undirected graph $G=(V,E)$ with $n=|V|$ vertices and $m=|E|$ edges, our algorithm identifies its connected components, and labels each vertex $v\in V$ with its corresponding component. Our algorithm works on an array of 3-tuples $\tuplE{p}{q}{r}$, where $p$, $q$, and $r$ are integers. The first two elements of these tuples will be updated in each iteration of the algorithm. The third element $r$ corresponds to a vertex $r \in V$ of the graph and is not changed throughout the algorithm. This element will also be used to identify the vertices of $G$ with their final connected components after termination.

Let $\mathcal{A}_i$ denote the array of tuples in iteration $i$. We initialize $\mathcal{A}_0$ as follows: for each vertex $x \in V$, we add the tuple $\tuplE{x}{\_}{x}$, and for each undirected edge $\{x,y\} \in E$, we add tuples $\tuplE{x}{\_}{y}$ and $\tuplE{y}{\_}{x}$. The middle elements will be initialized later during the algorithm. 

We denote the set of unique values in the first entry of all the tuples in $\mathcal{A}_i$ by $\mathcal{P}_i$, therefore $\mathcal{P}_i = \{p \mid \tuplE{p}{q}{r} \in \mathcal{A}_i\}$. We refer to the unique values in $\mathcal{P}_i$ as \emph{partitions}, which represent intermediate groupings of tuples that eventually coalesce into connected components. We say that a tuple $\tuplE{p}{q}{r}$ is a member of the partition $p$. Once the algorithm converges, all tuples for a vertex $r$ will have a single unique partition $p$, which is also the unique connected component label for this vertex. 

In order to refer to the tuples of a partition $p$, we define the \emph{partition bucket} $\mathcal{PB}_i(p)$ of $p$ as those tuples which contain $p$ in their first entry: $\mathcal{PB}_i(p) = \{\tuplE{\hat{p}}{q}{r} \in \mathcal{A}_i \mid \hat{p}=p\}$. Further, we define the \emph{candidates} or the next potential partitions $\mathcal{C}_i(p)$ of $p$ as the values contained in the second tuple position of the partition bucket for $p$:
$\mathcal{C}_i(p) = \{q \mid \tuplE{p}{q}{r} \in \mathcal{PB}_i(p)\}$. We denote the minimum of the candidates of $p$ as $p_{min} = \min \mathcal{C}_i(p)$. A partition $p$ for which $p_{min}=p$ is called a \emph{stable partition}. Further, to identify all the vertices in a partition, we define the \emph{vertex members} of a partition $p$ as $\mathcal{V}_i(p) = \{r \mid \tuplE{p}{q}{r} \in \mathcal{PB}_i(p)\}$.   

Each vertex $u \in V$ is associated with multiple tuples in $\mathcal{A}_i$, possibly in different partitions $p$. We define \emph{vertex bucket} $\mathcal{VB}_i(u)$ as those tuples which contain $u$ in their third entry: $\mathcal{VB}_i(u) = \{\tuplE{p}{q}{r} \in \mathcal{A}_i \mid r=u\}$. We define the partitions $\mathcal{M}_i(u)$ as the set of partitions in the vertex bucket for $u$: $\mathcal{M}_i(u) = \{p \mid \tuplE{p}{q}{r} \in \mathcal{VB}_i(u)\}$. The minimum partition in $\mathcal{M}_i(u)$, i.e., $\min \mathcal{M}_i(u)$ is called \emph{nominated partition} by $u$. 

For a small example graph with vertices $u, v_1, v_2$, (Fig. \ref{fig:algExample}), we show the  array of tuples $\mathcal{A}$. At the initialization stage, the vertex bucket $\mathcal{VB}_0(u)$ of $u$ is the set of tuples $\{ \tuplE{u}{\_}{u}, \tuplE{v_1}{\_}{u}, \tuplE{v_2}{\_}{u}  \}$. The set of unique partitions $\mathcal{P}_0$ equals $\{u, v_1, v_2\}$. The partition bucket $\mathcal{PB}_0(u)$ for partition $u$ is given by the set $\{ \tuplE{u}{\_}{u}, \tuplE{u}{\_}{v_1}, \tuplE{u}{\_}{v_2}  \}$. At termination of our algorithm,
all tuples will have the same common partition id, which for this example will be $\min (u, v_1, v_2)$.

Each partition is associated with a set of vertices, and the tuples
for a vertex can be part of multiple partitions. We define the neighborhood
for a partition $p$ as those partitions which share at least one vertex
with $p$, i.e., those which share tuples with a common identical value in the third tuple element.
More formally, we define the \emph{neighborhood} partitions of $p$ as $\mathcal{N}_i(p) = \cup_{u \in \mathcal{V}_i(p)}{\mathcal{M}_i(u)}$. In the above example, the neighborhood partitions $\mathcal{N}_0(v_1)$ for the partition $v_1$  are $u, v_1$ and $v_2$. All  the notations introduced in this section are summarized in Table \ref{tab:symbols} for quick reference.

 %Notations used in the algorithm

%----------
% Sequential algorithm
%----------
\subsubsection{Algorithm}\label{sec:algo_seq}
We first describe the sequential version of our algorithm, outlined in Algorithm \ref{alg:ccsortpd}. Our algorithm is structured similar to the classic \emph{Shiloach-Vishkin} algorithm.  However, our algorithm is implemented differently, using an edge-centric representation of the graph.  

At a high level, every vertex begins in its own partition, and partitions are connected via the edges of the graph. In each iteration, we join each partition to its numerically minimal neighbor, until the partitions converge into the connected components of the graph. In order to resolve large diameter components quickly, we utilize the pointer doubling technique during shortcutting. 
To implement pointer doubling, we will require the \emph{parent} partition id of the newly joined partition in each iteration. We use \emph{temporary} tuples $\tuplE{p}{q}{r}_{tmp}$ to fetch this information. These tuples will be created and erased within the same iteration.  

\begin{figure}[!t]
 	\removelatexerror
	\begin{algorithm}[H]
    	\caption{Connected components labeling}
        \label{alg:ccsortpd}
    	\KwIn{Undirected graph $G=(V,E)$}
        \KwOut{Labeling of Connected Components}
        $\mathcal{A}_0 = [ ] $ \\
       	\lFor{$x \in V$}{$\mathcal{A}_0$.append($\langle x, \_, x \rangle$)}
    	\lFor{$\{x,y\} \in E$}{$\mathcal{A}_0$.append($\langle x, \_, y \rangle$, $\langle y, \_, x \rangle$)}
        $i \leftarrow 1$ \\
        $converged \leftarrow $ false \\ 
        \While{$converged \neq true$ }
        {
        	$converged \leftarrow $ true \label{alg:loopFirstLine}\\
            $\mathcal{A}_i \leftarrow \mathcal{A}_{i-1}$ \\
        	$\mathcal{M}_i(u) \leftarrow$ \texttt{sort}$(\mathcal{A}_i$ by third element$)$ \label{alg:sort1}\\
            \For{$u \in V $} 
            {
                $u_{min} \leftarrow \min \mathcal{M}_i(u)$\\
                \For{\textbf{each } $\tuplE{p}{q}{r}  \in \mathcal{VB}_i(u)$}
                 {
                	$\langle p, \boldsymbol{q}, r \rangle \leftarrow \langle p, \boldsymbol{u_{min}}, r  \rangle$\\
                 }
            } \label{alg:endsort1}
        	$\mathcal{C}_i(p) \leftarrow$ \texttt{sort}$(\mathcal{A}_i$ by first element$)$ \label{alg:sort2}\\
            \For{$p \in \mathcal{P}_i$} 
            {
                $p_{min} \leftarrow \min \mathcal{C}_i(p)$\\
                \If{ $p \neq p_{min}$}   
                {										\label{alg:convergence}
                	$converged \leftarrow $ false\\
                }
                \For{\textbf{each } $\tuplE{p}{q}{r} \in \mathcal{PB}_i(p)$}
                 {
                	$\langle \boldsymbol{p},q, r \rangle \leftarrow \langle \boldsymbol{p_{min}}, q, r  \rangle$\\
                 }\label{alg:endsort2}
                 $\mathcal{A}_i$.append($\langle p_{min}, \_, p_{min} \rangle _{tmp}$)	\label{alg:appendTuples}
            }
            \textbf{redo} steps \ref{alg:sort1} - \ref{alg:endsort1} \label{alg:sort3} \\
            \textbf{redo} steps \ref{alg:sort2} - \ref{alg:endsort2} \label{alg:sort4} \\
            \For{\textbf{each } $\tuplE{p}{q}{r}  _{tmp} \in \mathcal{A}_i$ \label{algo:iterateTuples}} 
           	{
            	$\mathcal{A}_i$.erase($\tuplE{p}{q}{r}  _{tmp}$) \label{alg:removeTuples}
        	}
            $i \leftarrow i + 1$
        }
	\end{algorithm}
\caption{Our parallel SV algorithm, presented using sequential semantics.}
\end{figure} %Algorithm pseudo code

As laid out in Section \ref{sec:algo_defs}, we first create an array of tuples $\mathcal{A}$, containing one tuple per vertex and two tuples per edge (Algorithm \ref{alg:ccsortpd}). In each iteration $i$, we perform four sorting operations over $\mathcal{A}_i$. %by $r$ and $p$ of each $\tuplE{p}{q}{r}$ tuple in succession. 
During the first two sorting operations, we compute and join each partition $p$ to its minimum neighborhood, i.e. $\min \mathcal{N}_i(p)$. Sorting $\mathcal{A}_i$ by the third entry, namely the vertex ids enables easy and cache efficient processing of each vertex bucket $\mathcal{VB}_i(u), u \in V$, since the tuples of a bucket are positioned contiguously in $\mathcal{A}_i$ due to the sorted order  (line \ref{alg:sort1}-\ref{alg:endsort1}). For each vertex bucket $\mathcal{VB}_i(u)$, we scan all the partition ids containing $u$, i.e., $\mathcal{M}_i(u)$  and compute the nominated partition $u_{min}$ which becomes the candidate (potential next partition). We save the candidate partition id in the second element of the tuples.

After computing all the candidate partitions, we perform a second global sort of $\mathcal{A}_i$ by the first tuple element in order to process the partition buckets $\mathcal{PB}_i$ (line \ref{alg:sort2}-\ref{alg:endsort2}). Each partition $p \in \mathcal{P}_i$ then computes and joins the minimum candidate partition, i.e., $p_{min} = \min \mathcal{C}_i(p)$. 
In other words, partition $p$ joins its minimum neighbor $p_{min}$. We loop over these two sort-and-update steps until partitions converge into the connected components of the graph. Convergence for a partition $p$ is reached when its \emph{neighborhood} $\mathcal{N}_i(p)$ contains $p$ as its only member. 
Consequently, we can determine when to terminate the algorithm by checking whether all the partitions have fully converged, i.e., if they do not have any further neighboring partitions. For any partition $p$, $p \neq p_{min}$ implies the existence of at least one neighbor partition around $p$ (line \ref{alg:convergence}). 

\begin{figure}[!t]
\includegraphics[width=3.2in]{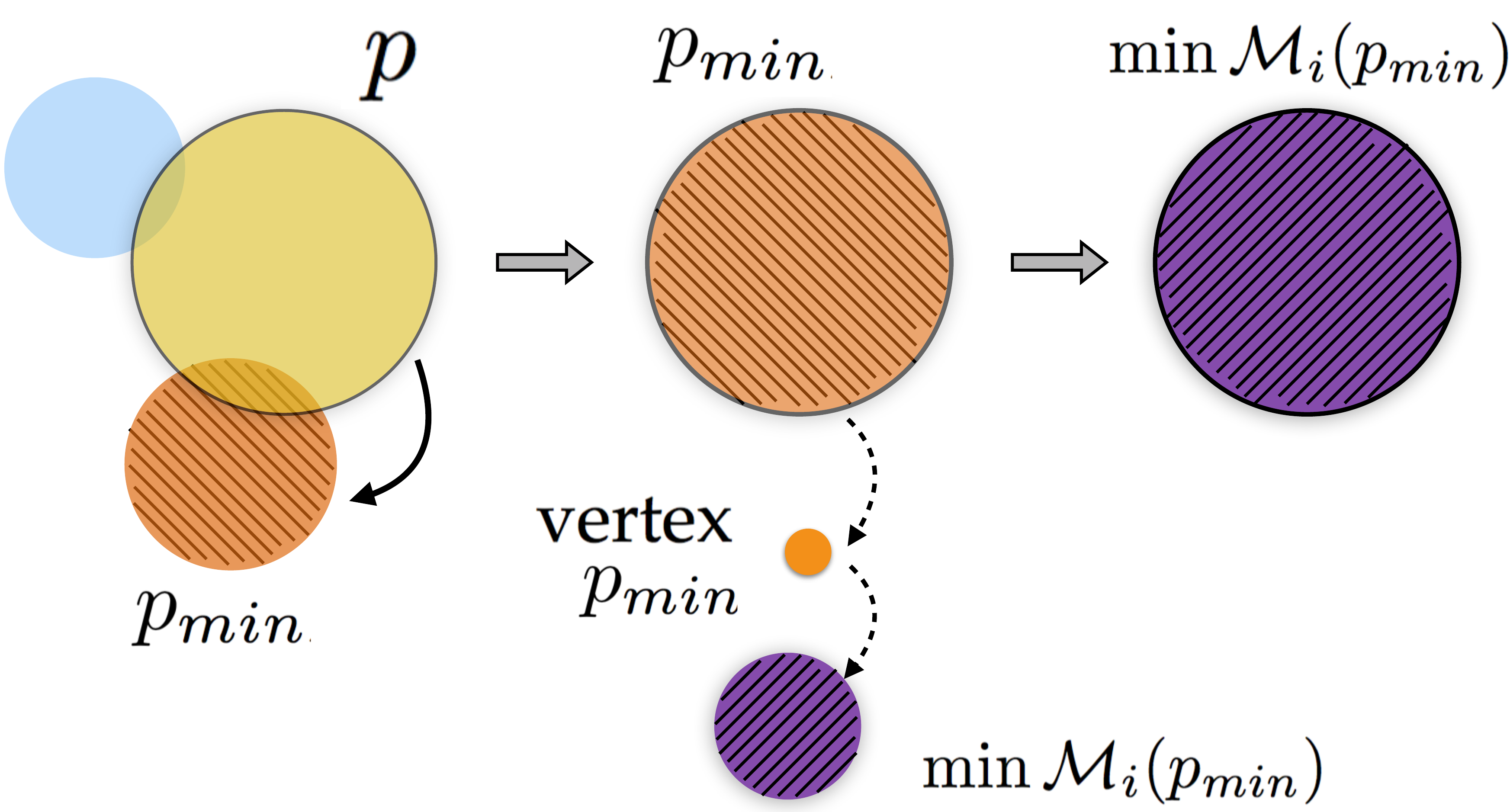}
\caption{Role of the four sorting phases used in each iteration of the algorithm. Using the first two sorts, partition $p$ joins $p_{min}$. The next two sorts enable pointer-jumping as $p_{min}$ joins $\min \mathcal{M}_i(p_{min})$. The temporary tuple $\tuplE{p_{min}}{\_}{p_{min}}_{tmp}$ used in the algorithm simulates a link between the partition $p_{min}$ and the vertex $p_{min}$ to allow jumping.} 
\label{fig:doubling}
\end{figure}

Iteratively invoking lines \ref{alg:loopFirstLine}-\ref{alg:endsort2} until convergence produces connected components of the graph within $O(n)$ iterations in the worst-case. By following the pointer doubling technique described in the SV algorithm~\cite{shiloach1982logn}, we achieve logarithmic convergence.  We summarize the role of all the four sorting operations in Figure \ref{fig:doubling}. After joining partition $p$ to $p_{min}$, we revise $p_{min}$ to $\min \mathcal{M}_i(p_{min})$.  The revision is effected by introducing temporary tuples $\tuplE{p_{min}}{\_}{p_{min}}_{tmp}$ in $\mathcal{A}_i$ (line \ref{alg:appendTuples}), then repeating the two sorts by the third and first element respectively (line \ref{alg:sort3}, \ref{alg:sort4}). In a way similar to the first two sorts of this iteration, the third sort forces the vertex $p_{min}$ to nominate $\min \mathcal{M}_i(p_{min})$ as the candidate partition id in the second element of the temporary tuples. Partition $p_{min}$, then, joins the partition id $\min \mathcal{M}_i(p_{min})$ after the final sort. The temporary tuples are removed from $\mathcal{A}_i$ after the pointer doubling phase is completed (line \ref{alg:removeTuples}). 

Note that the global count of the temporary tuples equals $|\mathcal{P}_i|$ in each iteration, and we know $|\mathcal{P}_i| \leq |V|$ (by the definition of $\mathcal{P}_i$). Therefore, the $O(m+n)$ bound holds for $|\mathcal{A}_i|$ throughout the execution. After the algorithm converges, the unique connected component label $c$ of a vertex $u \in V$ can be projected from the first element of any tuple $\tuplE{c}{\_}{u}$ in $\mathcal{A}$.   

\subsubsection{Parallel Algorithm}
\label{sec:distSVNaive}

We now describe our parallel implementation of the above algorithm for connected components labeling in a distributed memory environment. In this setting, each processor in the environment has its own locally addressable memory space. Remote data is accessible only through well defined communication primitives over the interconnection network.  The algorithm consists of three components: data distribution, parallel sorts, and bucket updates.  We designed our algorithm and its components using MPI primitives.

{\bf Data Distribution}:
All data, including the input, intermediate results, and final output, are equally distributed across all available processors. As specified in section \ref{sec:algo_seq}, the pipeline begins by generating tuples of the form $\tuplE{p}{q}{r}$ from the block distributed input $G(V,E)$ as edge list. By the end of this operation, each of the $\rho$ processes contains its equal share of $|\mathcal{A}|/\rho$ tuples.

{\bf Parallel Sorts}:
The bulk step of the algorithm is the sorting of tuples by either their third or first element in order to form the buckets $\mathcal{VB}_i$ or $\mathcal{PB}_i$, respectively. Parallel distributed memory sorting has been studied extensively. Blelloch {\it et al.} \cite{blelloch1991comparison} give a good review of different methods. With sufficiently large count of elements per process, which is often true while processing large datasets, the study concluded that samplesort is the fastest. Accordingly, we implement a variant of samplesort with regular sampling, where each processor first sorts its local array independently, and then picks equally spaced samples. The samples are then again sorted and $\rho-1$ of these samples are used as splitters for distributing data among processors. In a final step, the sorted sequences are merged locally.

{\bf Bucket Updates}:
After each sort, we need to determine the minimum element for each bucket, either $u_{min}$ for $\mathcal{VB}_i(u)$ or $p_{min}$ for $\mathcal{PB}_i(p)$. As a result of the parallel sorting, all the tuples $\tuplE{p}{q}{r}$ belonging to the same bucket are stored consecutively. However, a bucket might span multiple processors. 
Therefore, the first and last bucket of each processor require global communication during processing, while the internal buckets are processed in the same way as in the sequential case.  
Note, the first and last bucket on a processor may be the same if a bucket spans an entire processor. Communicating the minimum of buckets with the previous and next processor would require $O(\rho)$ communication steps in the worst case, since large $O(|\mathcal{A}|)$ size partitions can span across $O(\rho)$ processes. We thus use two parallel prefix (scan) operations with custom operators to achieve independence from the size of partitions, requiring at most $O(\log \rho)$ communication steps in addition to the local linear time processing time.

We describe the custom reduction operation to compute the $p_{min}$ within the partition buckets $\mathcal{PB}_i(p)$. Note that when computing $p_{min}$ in the algorithm, $\mathcal{A}_{i}$ is already sorted by the first element of the tuples and $p_{min}$ is the minimum second element for tuples in each bucket. We first perform an exclusive scan, where each processor participates with the minimum tuple from its last bucket. This operation communicates the minimum of buckets from lower processor rank to higher rank. The binary reduction operator chooses from 2 tuples the tuple $\tuplE{p}{q}{r}$ with the maximum $p$, and between those with equal $p$, the minimum $q$.  Next we perform a reverse exclusive prefix scan to communicate the minimum from high rank to low rank. Here, each processor participates with its minimum tuple of its first bucket. Given the two results of the scan operations, we can compute for each processor the overall minimum $p_{min}$ for both the first and the last buckets. Computing $u_{min}$ follows a similar procedure.

{\bf Runtime Complexity}:
The runtime complexity of each iteration is dominated by sorting $\mathcal{A}$, and the number of iterations is bounded by $O(\log n)$. If $T(k,\rho)$ is the runtime to sort $k$ elements using $\rho$ processes, the runtime of our algorithm for computing connectivity of graph $G(V,E)$ equals $O(log(n) \cdot T(m + n, \rho))$. %Because the algorithm is based on Shiloach-Vishkin's approach, it is not work-optimal. 

\subsubsection{Excluding Completed Partitions}\label{sec:distSVOpt1}
As the algorithm progresses through iterations, certain partitions become \emph{completed}. A partition $p$ is \emph{completed} if $p$ has no neighbor partition except itself, i.e., $\mathcal{N}_i(p) = \{p\}$. Even though we have described how to detect the global convergence of the algorithm, detecting as well as excluding the \emph{completed} partitions reduces the active working set throughout successive iterations. 

By the definition of $\mathcal{N}_i(p)$ in Section \ref{sec:algo_defs}, $\mathcal{N}_i(p) = \{p\}$ implies that $\cup_{u \in \mathcal{V}_i(p)} \mathcal{M}_i(u) = \{p\}$. Since the third elements of the tuples are never altered, each vertex is associated with at least one partition throughout the algorithm, therefore $|\mathcal{M}_i(u)| > 0 \,\forall u \in V$. Using these arguments, we claim the following: $p$ is \emph{completed} $\Leftrightarrow$ $\mathcal{M}_i(u) = \{p\} \,\forall u \in \mathcal{V}_i(p)$. Once the partition is \emph{completed}, it takes us one more iteration to detect its completion. While processing the vertex buckets after the first sort of the algorithm, we label all the tuples in $\mathcal{VB}_i(u) , u \in V$ as \emph{potentially completed} if $|\mathcal{M}_i(u)| = 1$. While processing the partition buckets subsequently, partition $p$ is marked as \emph{completed} if all the tuples in $\mathcal{PB}_i(p)$ are \emph{potentially completed}.

Completed partitions are marked as such and swapped to the end of the local array. All following iterations treat only the first, non-completed part of its local array as the local working set. As a result, the size of the active working set shrinks throughout successive iterations. This optimization yields significant reduction in the volume of active data, particularly for graphs with a large number of small components, since many small connected components are quickly identified and excluded from future processing.

\subsubsection{Load Balancing}\label{sec:distSVOpt2}
Although we initially start with a block decomposition of the array $\mathcal{A}$, exclusion of \emph{completed partitions} introduces an increasing imbalance of the active elements with each iteration. Since we join partitions from larger \emph{ids} to smaller \emph{ids}, a large partition will have smaller final partition ids than small partitions probabilistically. As the sort operation maps large id partitions to higher rank processes, the higher rank processes retain fewer and fewer active tuples over time, while lower rank processes contain growing partitions with small ids. Our experiments in Section \ref{sec:performance} study this imbalance of data distribution and its effect on the overall run time. We resolve this problem and further optimize our algorithm by evenly redistributing the active tuples after each iteration. Our results show that this optimization yields significant improvement in the total run time.

\subsection{Hybrid Implementation using BFS}\label{sec:bfs}
Connected components can be found using a series of BFS traversals, one for each component.
The known parallel BFS algorithms are asymptotically work-optimal, i.e., they maintain $O(m + n)$ parallel work for small-world networks~\cite{bulucc2011parallel}. 
Parallel BFS software can be adapted to achieve the same objective as our parallel SV algorithm, namely to compute all the connected components in a graph. To do so, parallel BFS can be executed iteratively, each time selecting a new seed vertex from among the vertices that were not visited during any of the prior BFS iterations. However, we note the following strengths and weaknesses associated with using BFS methods for the connectivity problem:

\begin{itemize}
\item \textbf{Pro:} For a massive connected component with a small diameter, the large number of vertices at each level of the traversal yields enough data parallelism for parallel BFS methods to become bandwidth bound, and thus efficient.  

\item \textbf{Con:} When the diameter of a component is large and vertex degrees are small, for instance in mesh graphs, the number of vertices at each level of BFS traversal is small.  The application becomes latency-bound due to the lack of data parallelism. This leads to under-utilization of the compute resources and the loss of efficiency in practice~\cite{bulucc2011parallel}.

\item \textbf{Con:} For graphs with a large number of small components, parallel BFS needs to be executed repeatedly. The application becomes latency-bound as the synchronization and remote communication latency costs predominate the effective work done during the execution. In this case, BFS method's scalability is greatly diminished. Slota {\it et al.}~\cite{slota2014bfs} draw a similar conclusion while parallelizing the strongly connected components problem using shared memory systems. 

\end{itemize}

A small world scale-free network contains a single large connected component \cite{easley2010networks}. To compute the connectivity of these graphs, we note that identifying the first connected component using a BFS traversal is more efficient than using the SV algorithm over the complete graph. For parallel BFS, we use Bulu\c{c} {\it et al.}'s \cite{bulucc2011parallel} state-of-the-art implementation available as part of the CombBLAS library \cite{bulucc2011combinatorial} and integrate this software as an alternative pre-processing step to our parallel SV algorithm.      

Scale-free networks are characterized by a power-law vertex degree distribution \cite{barabasi2000scale}. Therefore, we classify the graph structure as scale-free by checking if the degree distribution follows a power-law distribution. We use the statistical framework described by Clauset \textit{et al.} \cite{clauset2009power} to fit a power-law curve to the discrete graph degree distribution, and estimate the goodness of fit with one-sample Kolmogorov-Smirnov (K-S) test. The closer the K-S statistic value is to 0, the better is the fit.  If this value is below a user specified threshold $\tau$, then we execute a BFS iteration before invoking our parallel SV algorithm. Algorithm \ref{alg:cchybrid} gives the outline of our hybrid approach.

\begin{figure}[!t]
 	\removelatexerror
	\begin{algorithm}[H]
    	\caption{Connected components labeling}
        \label{alg:cchybrid}
    	\KwIn{Undirected graph $G=(V,E)$}
        \KwOut{Labeling of Connected Components}
        // Graph structure prediction \\
        $\mathcal{D} \leftarrow $ Degree\_Distribution$(G)$ \label{alg2:degdist} \\
        \If {K-S statistic ($\mathcal{D}$) $ < \tau$}{ \label{alg2:statistics} 
        	// Relabel vertices \\
            $G(V,E)$ $\leftarrow G(V,E)$ s.t. $ u \in [0, |V| - 1] \,\forall u \in V$ \label{alg2:relabel} \\
            //Execute BFS \\
            choose a seed $s \in V$ \\
            $\mathcal{VI} \leftarrow $Parallel-BFS$(s)$ \\
            
            //Filter out the traversed component \\
            $V \leftarrow V \setminus \mathcal{VI}$ \label{alg2:filtervertices}\\
            $E \leftarrow E \setminus \{(u,v) | u \in \mathcal{VI}\}$ \label{alg2:filteredges} \\
        }
        Parallel-SV$(G(V,E))$ \label{alg2:runSV}\\
	\end{algorithm}
\caption{Hybrid approach using parallel BFS and SV algorithms to compute connected components}
\vspace*{-1cm}
\end{figure}

In our implementation, we choose to store each undirected edge $(u,v)$ as two directed edges $(v,u)$ and $(u,v)$ in our edge list. This simplifies the computation of the degree distribution of the graph (line \ref{alg2:degdist}). We compute the degree distribution $\mathcal{D}$ of the graph by doing a global sort of edge list by the source vertex. Through a linear scan over the sorted edge list, we compute the degree of each vertex $u \in V$. In practice, it is safe to assume that the maximum vertex degree $c$ is much smaller than number of edges $|E|$ ($c \ll |E|$). Thus each process can compute the local degree distribution in an array of size $c$, and a parallel reduction operation is used to solve for $\mathcal{D}$. Once $\mathcal{D}$ is known, evaluating the degree distribution statistics takes insignificant time as size of $\mathcal{D}$ equals $c$. Therefore, we compute the K-S statistics as described before, sequentially on each process.      

If the K-S statistic is below the set threshold, we choose to run the parallel BFS on $G(V,E)$ (line \ref{alg2:statistics}). Bulu\c{c}'s BFS implementation works with the graph in an adjacency matrix format. Accordingly, we relabel the vertices in $G(V,E)$ such that vertex ids are between $0$ to $|V|-1$ (line \ref{alg2:relabel}). This process requires sorting the edge list twice, once by the source vertices and second by the destination vertices. After the first sort, we perform a parallel prefix (scan) operation to label the source vertices with a unique id $\in [0, V-1]$. Similarly, we update the destination vertices using the second sort.

Next, we execute the parallel BFS from a randomly selected vertex in $G(V,E)$ and get a distributed list of visited vertices $\mathcal{VI}$ as the result. Note that the visited graph component is expected to be the largest one as it spans the majority of $G(V,E)$ in the case of scale-free graphs. To continue solving for other components, we filter out the visited component $\mathcal{VI}$ from $G(V,E)$ (line \ref{alg2:filtervertices},\ref{alg2:filteredges}). $\mathcal{VI}$ is distributed identically as $V$, therefore vertex filtering is done locally on each process. We already have the edge list $E$ in the sorted order by destination vertices due to the previous operations, therefore we execute an all-to-all collective operation to distribute $\mathcal{VI}$ based on the sorted order and delete the visited edges locally on each processor. Finally, irrespective of whether we use BFS or not, we run the parallel SV algorithm on $G(V,E)$ (line \ref{alg2:runSV}). In our experiments, we show the overall gain in performance using the hybrid approach as well as the additional overhead incurred by the prediction phase. We also report the proportion of time spent in each of the prediction, relabeling, parallel-BFS, filtering, and parallel-SV stages.

\section{Experimental Setup}
%-----------
%----TESTBED
%-----------
\subsection*{Hardware}
For the experiments, we use Edison, a Cray XC30 supercomputer located at Lawrence Berkeley National Laboratory. In this system, each of the 5,576 compute nodes has two 12-core Intel Ivy Bridge processors with 2.4 GHz clock speed and 64 GB DDR3 memory. To perform parallel I/O, we use the scratch storage supported through the Lustre file system. We assign one MPI process per physical core for the execution of our algorithm. Further, we only use square process grids as CombBLAS \cite{bulucc2011combinatorial} requires the process count to be a perfect square.
%-----------
%----DATASETS
%-----------
\subsection*{Datasets}
\begin{table*}[!t]
\renewcommand{\arraystretch}{1.3}
\caption{List of the nine graphs and their sizes used for conducting experiments. Edge between two vertices is counted once while reporting the graph sizes. Largest component's size is computed in terms of percentage of count of edges in the largest component relative to complete graph.}
\centering
\begin{tabular}{|p{0.5cm}|p{1.4cm}|p{1.6cm}|p{1cm}|p{1.4cm}|p{1.6cm}|p{1.2cm}|p{1.5cm}|p{1.6cm}|} \hline
\bfseries Id	&	\bfseries Dataset& \bfseries Type 	&\bfseries Vertices &\bfseries Undirected Edges & \bfseries Components & \bfseries Approx. diameter & \bfseries Largest component & \bfseries Source\\ \hline\hline

M1 			&Lake Lanier			&Metagenomic	&	1.1 B	&	1.1 B	& 2.6 M		& 3,763	& 53\%		& NCBI (SRR947737)		\\ \hline
M2 			&Human Metagenome 		&Metagenomic	&	2.0 B	&	2.0 B	& 1.0 M		& 3,989 & 91.1\%	& NCBI (SRR1804155)		\\ \hline
M3 			&Soil (Peru)			&Metagenomic	&	531.2 M	&	523.6 M	& 7.6 M		& 2,463 & 0.3\%		& MG-RAST (4477807.3)				\\ \hline
M4 			&Soil (Iowa)			&Metagenomic	&	53.7 B	&	53.6 B	& 319.2 M	& -		& 44.2\%	& JGI (402461)				\\ \hline
G1 			&Twitter				&Social			&	52.6 M	&	2.0 B	& 29,533	& 16	& 99.99\%	& \cite{cha2010measuring}					\\ \hline
G2 			&sk-2005				&Web Crawl		&	50.6 M	&	1.9 B	& 45		& 27	& 99.99\%	& \cite{davis2011university} 					\\ \hline
G3 			&eu-usa-osm				&Road Networks	&	74.9 M	&	82.9 M	& 2			& 25,105& 65.2\%	& \cite{davis2011university} 					\\ \hline
K1 			&Kronecker (scale = 27) &Kronecker		&	63.7 M	&	2.1 B	& 19,753	& 9		& 99.99\%	& Synthetic \cite{murphy2010introducing}					\\ \hline
K2 			&Kronecker (scale = 29) &Kronecker		&	235.4 M	&	8.6 B	& 73,182	& 9		& 99.99\%	& Synthetic \cite{murphy2010introducing} 					\\ \hline
\end{tabular}
\label{tab:data}
\end{table*}
Table \ref{tab:data} lists the 9 graphs used in our experiments. These include 4 de Bruijn graphs constructed from different metagenomic sequence datasets, one social graph from Twitter, one web crawl, one road network and two synthetic Kronecker graphs from the Graph500 benchmark. The sizes of these graphs range from 83 million edges to 54 billion edges. 

For each graph, we report the relevant statistics in Table \ref{tab:data} to correlate them with our performance results. Computing the exact diameter is computationally expensive and often infeasible for large graphs \cite{shun2015evaluation}. As such, we compute their approximate diameters by executing a total of 100 BFS runs from a set of random seed vertices. For all the graphs but M4, this approach was able to give us an approximation. However, the size of M4 required a substantial amount of time for completing this task and as such it did not complete. We estimate that only 4 of the 9 tested graphs are small world networks.   

\subsubsection*{Metagenomic de Bruijn Graphs} 
M1-M4 are built using publicly available metagenomics samples from different environments. We obtained the sequences in FASTQ format. We discarded the sequences with unknown nucleotides using the fastx\_clipper utility supported in the FASTX toolkit \cite{gordon2010fastx}. 
The size of the sequence dataset depends upon the amount of sampling done for each environment. We build de Bruijn graphs from these samples using the routines from the parallel distributed memory $k$-mer indexing library Kmerind~\cite{tonykmerind}. It is worth noting that in de Bruijn graphs, vertex degrees are bounded by $8$~\cite{compeau2011apply}.
One motivation for picking samples from different environments is the difference in graph properties associated with them such as the number of components and relative sizes. These are dependent on the degree of microbial diversity in the environments. Among the environments we picked, it has been estimated that the soil environments are the most diverse while the human microbiome samples are the least diverse of these environments \cite{rodriguez2014estimating}. This translates to large number of connected components in the soil graphs M3 and M4.

\subsubsection*{Other Graphs}
Graphs K1-K2 and G1-G3 are derived from widely used graph databases and benchmarks. We use the synthetic Kronecker graph generator from the Graph500 benchmark specifications \cite{murphy2010introducing} to build Kronecker graphs with scale 27 (K1) and 29 (K2). Graphs G1-G3 are downloaded directly from online databases in the edge list format. G1 and G2 are small world scale-free networks from twitter and online web crawl respectively. G3 consists of two road networks from Europe and USA, downloaded from the Florida Sparse Matrix Collection \cite{davis2011university}. Among all our graphs, G3 has the highest estimated diameter of 25K. To read these data files in our program, a file is partitioned into equal-sized blocks, one per MPI process.  The MPI processes concurrently read the blocks from the file system and generate distributed arrays of graph edges in a streaming fashion.

\section{Performance Analysis}\label{sec:performance}
In all our experiments, we exclude file I/O and de Bruijn graph construction time from our benchmarks, and begin profiling after the block-distributed list of edges are loaded into memory. Profiling terminates after computing the connected component labels for all the vertices in the graph. Each vertex id in the input edge list is assumed to be a 64 bit integer. The algorithm avoids any runtime bias on vertex naming of the graph by permuting the vertex ids using Robert Jenkin's 64 bit mix invertible hash function \cite{jenkins1997hash}.

\subsection*{Load Balancing}
\begin{figure}[!t]
\includegraphics[width=3.4in]{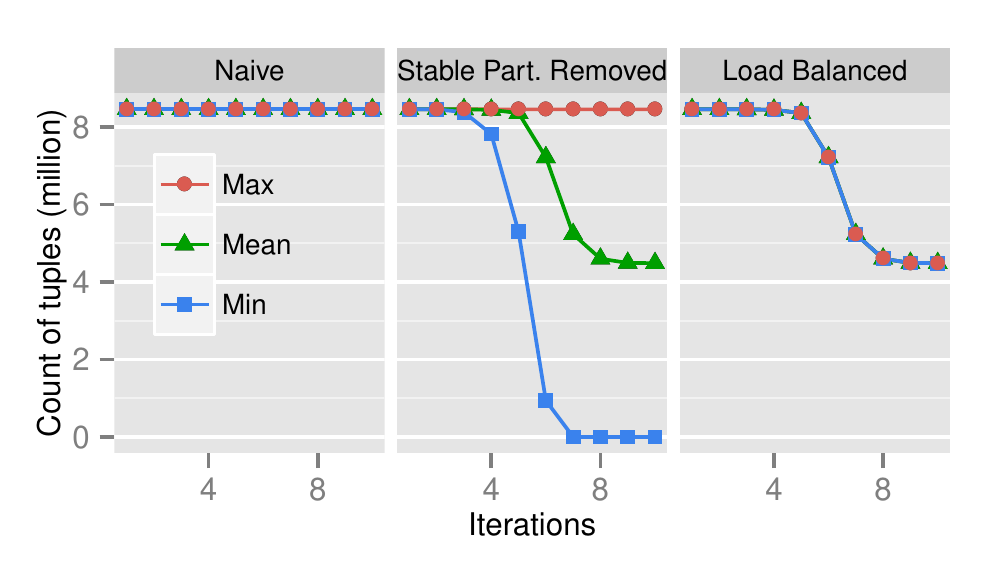}
\caption{Work load balance in terms of tuples per processes during each iteration of the three algorithm variants for parallel SV algorithm. 
Illustrated are the maximum, average, and minimum count of tuples on all the processes.
The experiments were conducted using the M1 graph and 256 processor cores. Each edge is represented as 2 tuples internally in the algorithm.}
\label{fig:loadbal}
\end{figure}

We first show the impact of the two optimizations performed by our parallel SV algorithm (Sections \ref{sec:distSVOpt1}, \ref{sec:distSVOpt2}) for reducing and balancing the work among the processes. Our algorithm used 10 iterations to compute the connectivity of M1. Figure \ref{fig:loadbal} shows the minimum (min), maximum (max), and mean size of the distributed tuple array per process as iterations progress in three variants of our algorithm, using 256 cores. The max load is important as it determines the parallel runtime. A smaller separation between the min and max values indicates better load balance. The first implementation, referred to as Naive (Section \ref{sec:distSVNaive}), does not remove the completed components along the iterations and therefore the work load remains constant. Removing the stable components reduces the size of the working set per each iteration as illustrated by the desirable decrease in mean tuple count. 
The difference between min and max grows significantly after 4 iterations. With our load balanced implementation, we see an even distribution of tuples across processors, as the minimum and maximum count are the same for each iteration. We see that the mean drops to about 50\% of the initial value because the largest component in M1 contains 53\% of the total edges (Table \ref{tab:data}). 

\begin{figure}[!t]
\includegraphics[width=3.4in]{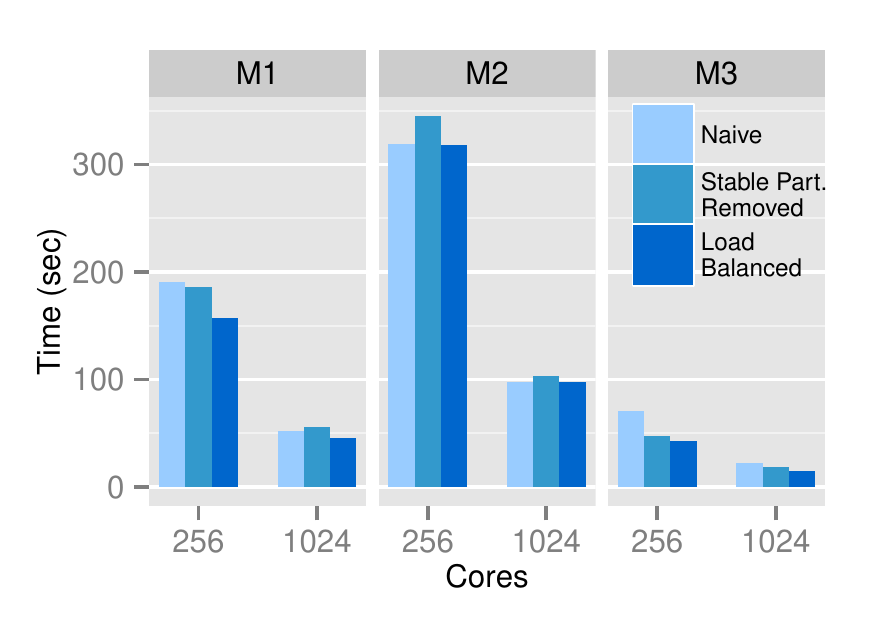}
\caption{Performance gains due to load balancing for graphs M1-M3 using 256, 1024 processor cores.}
\label{fig:loadbalTime}
\end{figure}

Consequently, we see improvement in the execution time for M1 and M3 in Figure \ref{fig:loadbalTime} as a result of these optimizations. Of the three implementations, the load balanced implementation consistently achieves better performance against the other two approaches. 
For the M2 graph, we get negligible gains using our load-balanced approach against the Naive approach because the largest component in M2 covers 91\% of the graph. Therefore, the total work load stays roughly the same across the iterations. 

% K-S statistics
\subsection*{Hybrid Implementation Analysis}
As discussed in Section \ref{sec:bfs}, BFS is more efficient for computing the first component in the small world scale-free graphs. We use an open-source C++ library \cite{ntamasplfit} which fits the power-law distributions to discrete empirical data based on the procedure described by Clauset \textit{et al.} \cite{clauset2009power}.  Table \ref{tab:ks_stat} shows the K-S statistic value computed using the degree distribution for all our graphs. For each of the graphs with scale-free topology (G1, G2, K1, K2), there is a clear distinction of these values against rest of the graphs. Based on these observations, we set a threshold of 0.05 to predict the scale-free structure of the underlying graph topology and execute a BFS iteration for such cases.

\begin{table}[!t]
\renewcommand{\arraystretch}{1.3}
\caption{Kolmogorov Smirnov test values used to estimate the goodness of power law curve fit to the degree distribution of each graph. BFS is executed if K-S statistic value is less than $0.05$.}
\centering
\begin{tabular}{|p{1.0cm}|p{1.4cm}|p{1.3cm}|p{1.3cm}|} \hline
\bfseries Dataset	&	\bfseries K-S statistic		& \bfseries Run BFS iteration? & \bfseries Correct Decision\\ \hline\hline
M1 			&	$0.41$		&	\xmark	&	\cmark	\\ \hline
M2 			&	$0.24$		&	\xmark	&	\xmark	\\ \hline	
M3 			&	$0.39$		&	\xmark	&	\cmark  \\ \hline
M4 			&	$0.31$		&	\xmark	&	\cmark  \\ \hline
G1 			&	$\bf{0.01}$	&	\cmark	&	\cmark 	\\ \hline
G2 			&	$\bf{0.03}$	&	\cmark	&	\cmark 	\\ \hline
G3 			&	$0.21$		&	\xmark	&	\cmark  \\ \hline
K1 			&	$\bf{0.01}$	&	\cmark	&	\cmark 	\\ \hline
K2 			&	$\bf{0.01}$	&	\cmark	&	\cmark 	\\ \hline
\end{tabular}
\label{tab:ks_stat}
\end{table}

\begin{figure}[!t]
\subfloat[Comparison of runtime of our algorithm making decision to run BFS dynamically versus the implementation which is hard-coded to make the opposite decision for all the graphs, using 2025 processor cores. For all the graphs but M2, our decision is correct (Table \ref{tab:ks_stat}).]{
	\includegraphics[width=3.2in]{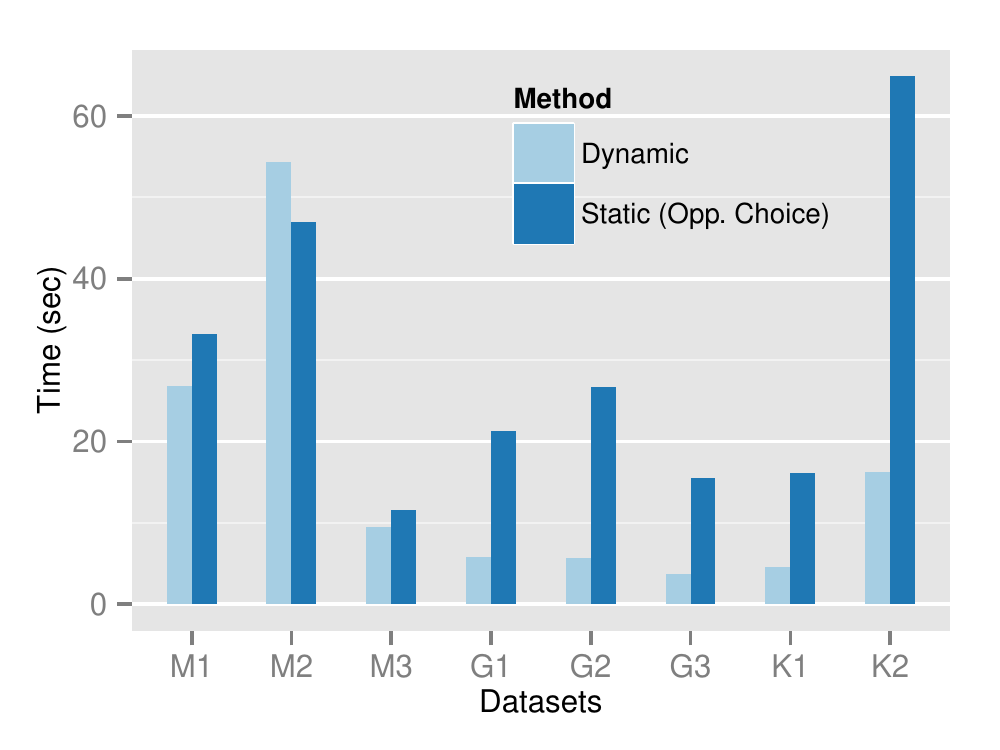}
    \label{fig:dynamic_opp}
}

\subfloat[Comparison of runtime of our algorithm making decision to run BFS dynamically versus the implementation hard-coded to make the same decision for all the graphs, using 2025 processor cores. The difference in the timings is the overhead of our prediction strategy.]{
	\includegraphics[width=3.2in]{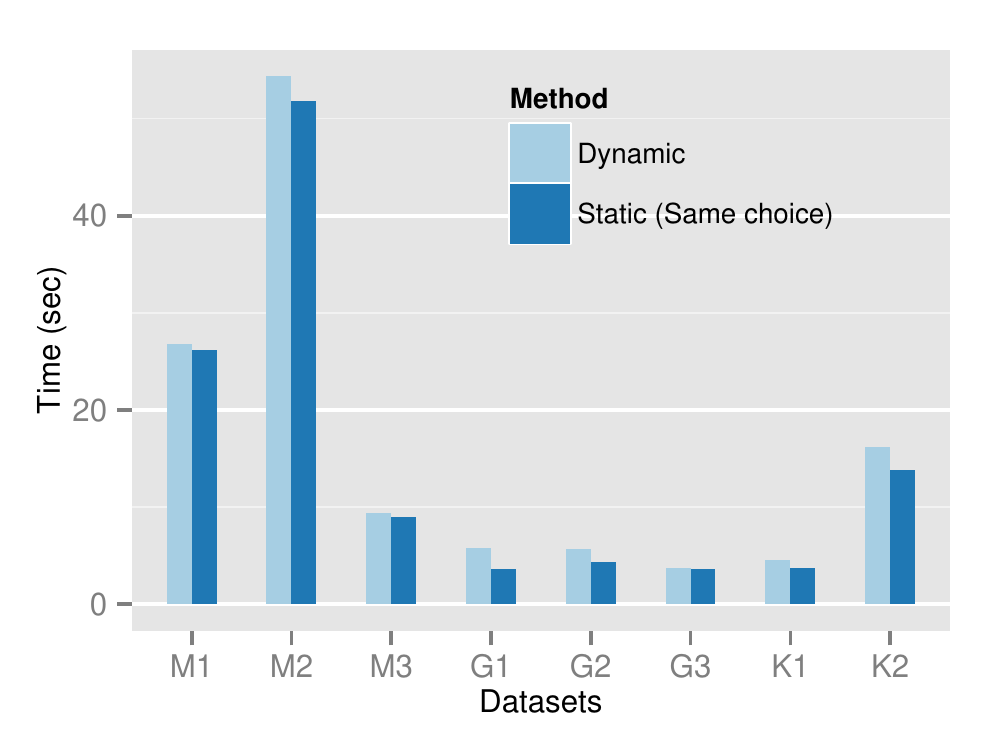}
    \label{fig:dynamic_same}
}
\caption{Evaluation of prediction heuristics in our algorithm}
\label{fig:dynamic}
\end{figure}

To measure the relative improvement obtained by running BFS iteration based on the prediction, we compare the runtime of this dynamic approach against our implementation that does not compute K-S statistics and is hard-coded to make the opposite choice, i.e., executing BFS iteration only for the graphs M1-M4, G3. This experiment, using 2025 processor cores, measures whether the prediction is correct and if correct, how much performance benefit do we gain against the opposite choice. As illustrated in figure \ref{fig:dynamic_opp}, we see positive speedups for all the graphs except M2. We see more than 3x performance gains for all the small world graphs as well as G3. For M1 and M3, we gained approximately 25\% improvement in the runtime. This experiment confirms that using BFS to identify and exclude the largest  component is much more effective for small world graphs while running BFS on large diameter graph such as G3 is not optimal. Moreover, using the degree distribution statistics, we can choose an optimal strategy for most of the graphs.   

Note that computing the degree distribution of a graph and measuring K-S statistics adds an extra overhead to the overall runtime of the algorithm. We evaluate the additional overhead incurred by comparing the dynamic approach against the implementation which is hard-coded to make the same choice, i.e., execute BFS iteration only for G1-G2, K1-K2 (Figure \ref{fig:dynamic_same}) using 2025 processor cores. The overhead varies from 60\% for G1 to only 2\% for M1. In general, we find this overhead to be relatively high for small-world graphs. Fitting the degree distribution curve against a power-law model is a sequential routine in our implementation, and it takes us about a second for scale-free graphs because they tend to have long-tailed degree distributions. We leave parallelizing and optimizing this routine as future work. Overall, we observe that the performance gains significantly outweigh the cost of computing the degree distribution and 
K-S test. 

% Scalability results, performance comparisons
\subsection*{Strong Scaling}
\begin{figure}[!t]
\includegraphics[width=3.2in]{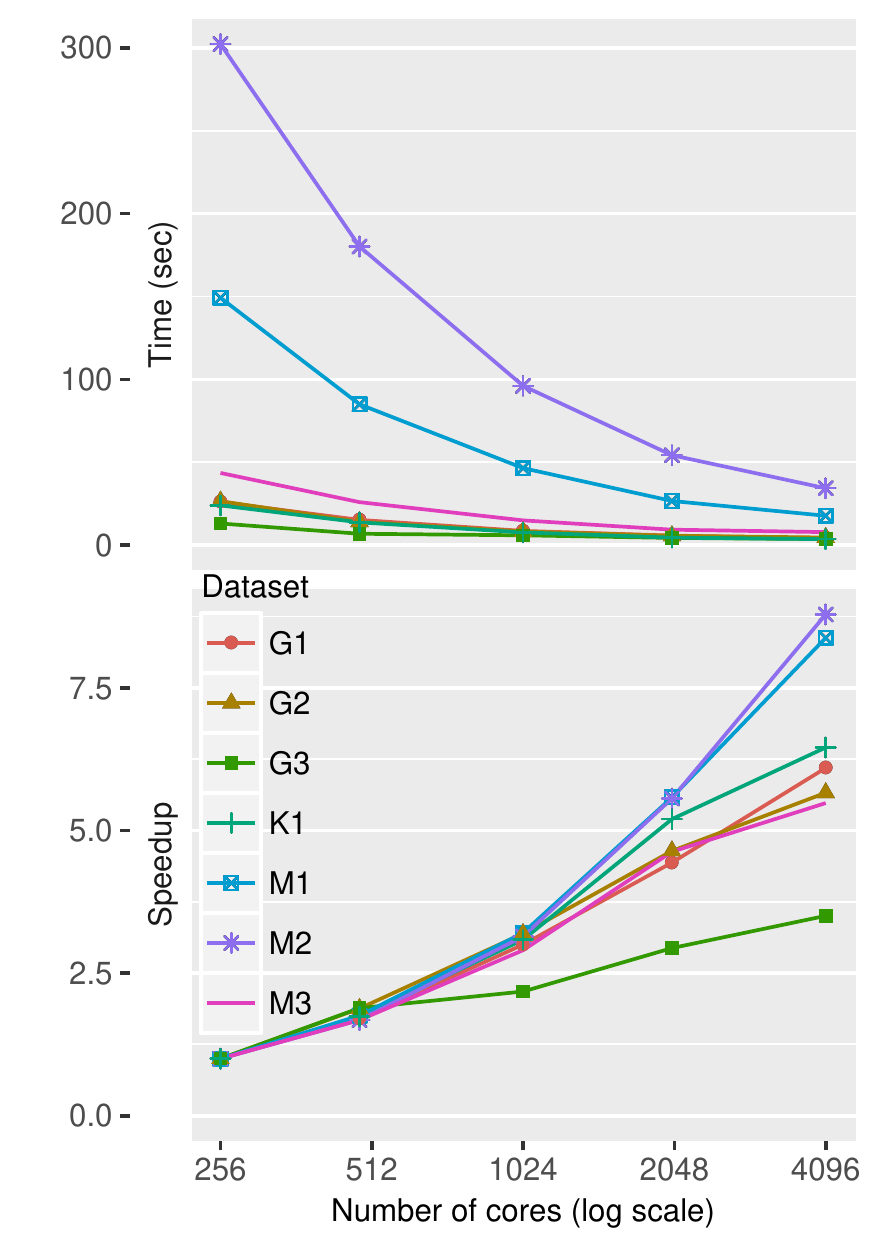}
\caption{Strong scalability results of our algorithm on different graphs using 4096 cores. Speedups are computed relative to the runtime on 256 cores.}
\label{fig:strongscale}
\end{figure}

With the optimizations in place, we conducted strong scaling experiments on our algorithm. In this experiment, we use 256-4096 cores for G1-G3, K1, and M1-M3. Results for M4, the largest graph are discussed separately as we could not process it with fewer than 4096 cores. Graph K2 is ignored for this experiment because it has same topology as K1. In Fig. \ref{fig:strongscale}, we show the runtimes as well as speedups achieved by our algorithm. Most of these graphs cannot fit in the memory of a single node, therefore speedups are measured relative to the runtime on 256 cores. Ideal relative speedup on 4096 cores is 16. We achieve maximum speedup of more than 8x for the metagenomic graphs M1 and M2 and close to 6x speedup for small world graphs G1, G2 and K1. G3 shows limited scalability due to its much smaller size relative to other graphs. We are able to compute connectivity for our largest graph M4 in 215 seconds using 32761 cores (Table \ref{tab:M4_timings}). 

In section \ref{sec:distSVNaive} we discussed how each iteration of our parallel SV algorithm uses parallel sorting to update the partition ids of the edges. As a majority of time of this algorithm is spent in performing sorting, we also execute a micro benchmark that sorts 2 billion randomly generated 64 bit integers using 256 and 4096 cores. Interestingly, we achieve speedup of 8.06 using our sample sorting method which is close to our scalability for M1 and M2. We anticipate that implementing more advanced sorting algorithms \cite{SK_IPDPS_2010} may further improve the efficiency of our parallel SV algorithm.

\begin{table}[!t]
\renewcommand{\arraystretch}{1.3}
\caption{Timings for the largest graph M4 with increasing processor cores}
\centering
\begin{tabular}{|p{2.4cm}|p{1.2cm}|p{1.2cm}|p{1.2cm}|} \hline
\bfseries Cores	& 8281 		& 16384 	&	32761		\\ \hline\hline
Time for M4 (sec) 	& 429.89	& 291.19 	& 	214.56		\\ \hline
\end{tabular}
\label{tab:M4_timings}
\end{table}

\subsection*{Performance Anatomy}
\begin{figure}[!t]
\includegraphics[width=3.5in]{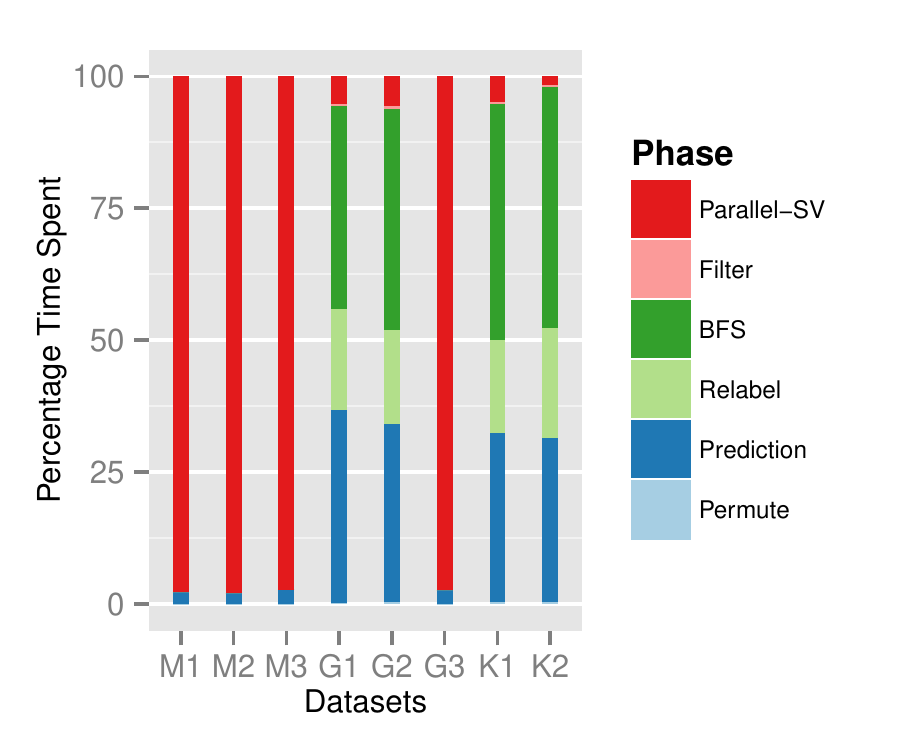}
\caption{Percentage time spend in different stages by the algorithm for different graphs using 2025 cores. BFS is executed only for graphs G1, G2, K1 and K2. }
\label{fig:breakdown}
\end{figure}

We also report the percentage of total execution time on 2025 cores that are attributable to each stage of our algorithm (Fig. \ref{fig:breakdown}). This figure is noteworthy especially for the graphs for which our algorithm chooses to execute BFS. For G1, G2, K1 and K2, more than 50\% of the total percentage of time is devoted to predicting the graph structure and relabeling the vertices before running the parallel BFS and SV algorithm. This figure is not meant to convey the true overhead due to the relabel and prediction operations individually, as the time for relabeling is reduced after we sort the edges during the prediction stage (Section \ref{sec:bfs}). Further, we measure the percentage time spent in the sorting operations in our parallel-SV algorithm for the graphs M1, M2, M3 and G3. As we expected, this measure is high and ranges from 91\% - 94\% for all the four graphs. 
  
\subsection*{Comparison with Previous Work}
\begin{figure}[!t]
\includegraphics[width=3.2in]{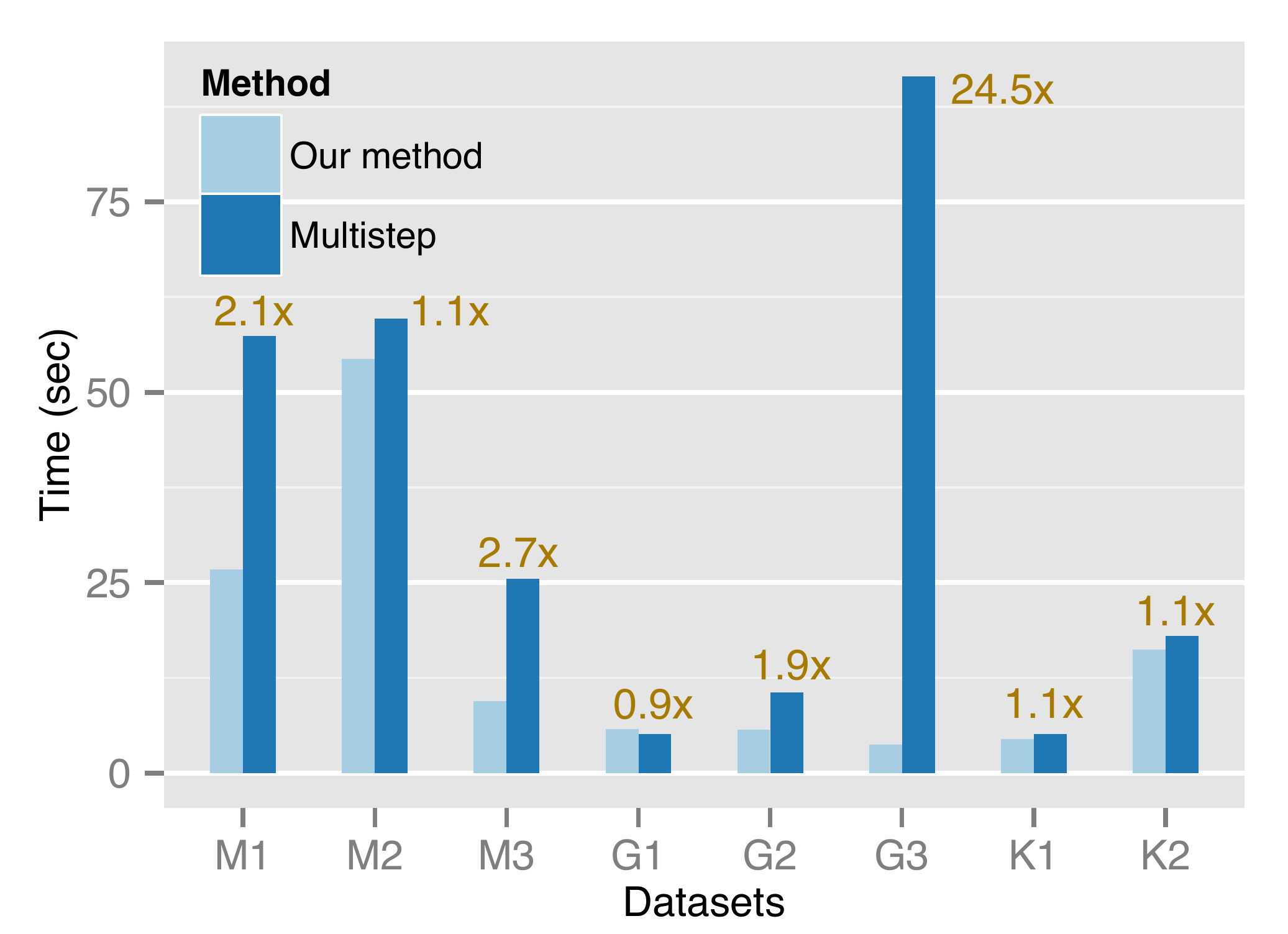}
\caption{Performance comparison of our algorithm against the Multistep method \cite{slota2016high} using multiple graphs with 2025 cores.}
\label{fig:compare}
\end{figure}

We achieve notable speedups when the performance of our algorithm is compared against the state-of-the-art Multistep algorithm \cite{slota2016high} using 2025 cores. As before, we begin counting the time once the graph edge list is read into the memory in both cases. We ran the Multistep algorithm with one process per physical core as we observed better performance doing so than using hybrid MPI-OpenMP mode. Also, because the Multistep method expects the vertex ids to be in the range 0 to $|V|-1$, we inserted our vertex relabeling routine in their implementation in order to run the software. Figure \ref{fig:compare} shows the comparison of our approach against the MultiStep method. We see $> 1$ speedups for our method in all the graphs except G1. The speedup achieved ranges from 1.1x for K2 to 24.5x for G3. The speedup roughly correlates with the diameter of the graphs. The improvements achieved for the graphs M1, M2, M3 and G1 can be attributed to two shortcomings in the Multistep approach: 1) It executes BFS for computing the first component in all the graphs. BFS attains limited parallelism for large diameter graphs due to small frontier sizes. 2) It uses the label propagation technique to compute other components which in the worst case can take as many iterations as the diameter of the graph to reach the solution.  

We could not compare our approach against the  distributed-memory graph contraction algorithm \cite{iverson2015evaluation} proposed to solve the connectivity problem, as the implementation is not open-source. Based on their experiment description, the graph contraction algorithm showed strong scalability only till 32 cores. Other distributed graph frameworks such as GraphX \cite{gonzalez2014graphx}, and FlashGraph \cite{zheng2015flashgraph} based on in-memory Apache Spark and external-memory framework, respectively, can compute the connectivity of large-scale graphs as well. Slota {\it el al.} \cite{slota2016high} show that their Multistep algorithm achieves superior performance against both of these methods. Because  our algorithm performs better than Multistep, we skip a direct comparison against GraphX and FlashGraph.  

\subsection*{Comparison with Sequential Implementation}
We examine the performance of our algorithm against the best known sequential implementation for computing connectivity, for graph instances which can fit in the single node memory (64 GB) - these are relatively small. Previous works \cite{patwary2012multi,patwary2010experiments} have shown that the Rem's method \cite{dijkstra1976discipline} based on the union-find approach achieves the best sequential performance. The sequential implementation we use in this algorithm was obtained from the authors of \cite{patwary2012multi}. Again, because the disjoint-set structure used in the algorithm requires the vertices to be numbered from $0$ to $n-1$, we placed our relabeling routine in the implementation. This experiment uses graphs M3 and G3, as all the other graphs require more than 64 GB memory. We also add a Kronecker graph of scale 25 ($m=537$M, $n=17$M) to include a short diameter graph instance. Results of this experiment are shown in Table \ref{tab:seq}. For these three graphs, our algorithm selects BFS iteration for Kronecker graph only. For the Kronecker graph, our algorithm achieves a 100x speedup using 1024 cores. For the other graphs, M3 and G3, where the SV algorithm is selected, the speedup decreases with respect to the sequential algorithm - which is partially due to the fact that the algorithm is not work optimal. 

\subsection*{Comparison with Shared-memory Implementations}
The objective of the following comparative discussion between the distributed and shared-memory algorithms is not only to discuss the performance difference where shared-memory implementations tend to get good scaling per core, rather it is to highlight some of the constraints that shared-memory implementations have in contrast to their distributed counterparts.

Shared-memory parallel methods \cite{patwary2012multi,shun2014simple} exhibit good speedups over the best sequential implementation. It is therefore of no surprise to us that these algorithms can outperform our algorithm, especially for small to mid-range graphs. However, there are numerous problem scales that these shared memory algorithms cannot cope with due to the size of the graph; whereas our algorithm can easily deal with such networks. Our parallel algorithm utilizes bulk synchronous communication instead of the fast asynchronous communication found in shared-memory frameworks. While such communications are inherently slower, they do enable processing larger networks.
Consider the largest network analyzed in this paper (Table \ref{tab:data}): metagenomic graph M4 which has $53.6$ billion edges and an equal number of vertices. Processing this graph in memory requires at least the following amount of memory: $2 \cdot (|V| + |E|)\times 8$ bytes. This assumes that the graph requires $|V|$ elements for the vertices and $2\cdot |E|$ elements for the edges \footnote{Recall that these are un-directed edges and it is customary in CSR, Compressed Sparse Row, format to store both directions of the edge}. Also, $|V|$ integers are required for tracking the connected component labels. Given the size of the graph, $4$ byte integers are not large enough to store all the unique keys and as such this requires using $8$ byte integers. For the M4 network, a total of $1.7$ TB DRAM is needed. As the sequencing cost continues to decline much faster than Moore's law \cite{muir2016real}, we envision the need to analyze even larger metagenomic graphs that require even more memory, in the near future. 
The problems of optimizing distributed-memory parallel algorithms while trying to attain peak performance continues to be an important challenge and one that deserves additional attention, especially the ability to reduce the overhead of communication.
 
\begin{table}[!t]
\renewcommand{\arraystretch}{1.3}
\caption{Performance comparison against Rem's sequential connectivity algorithm \cite{patwary2010experiments,dijkstra1976discipline} using 1024 cores.}
\centering
\begin{tabular}{|p{1.4cm}|p{1.8cm}||p{0.9cm}|p{0.9cm}|p{0.9cm}|} \hline
\multirow{2}{*}{\bfseries Dataset} &	\multirow{2}{*}{\parbox{1.8cm}{\bfseries Fastest Seq. Time (s)}}	& \multicolumn{3}{|c|}{\bfseries Speedup}\\
& & p = 64 & 256 & 1024 \\\hline\hline
Kronecker (25) & 228.8 & 10.1 & 34.3 & 100.6 \\\hline
M3 & 406.2 & 2.5 & 9.3 & 27.0 \\\hline
G3 & 45.9 & 0.9 & 3.5 & 7.6 \\\hline
\end{tabular}
\label{tab:seq}
\end{table}

% Shared-memory parallel methods \cite{patwary2012multi,shun2014simple} exhibit good speedups over the best sequential implementation, therefore, we expect these methods to outperform our algorithm by a similar order of magnitude. Note that our algorithm is designed and engineered to work efficiently for large massive graphs that do not fit in  the memory of single node. Our parallel algorithm does bulk synchronous communication instead of the fast asynchronous  communication possible in the parallel shared-memory frameworks. Note that computing connectivity of our largest metagenomic graph M4 requires at least $2(|V| + |E|)$ $8$-byte integers in memory including $|V|$ integers for tracking the connected component labels of each vertex and $|V| + 2|E|$ integers for storing the undirected graph in the Compressed Sparse Row format, which adds up to $1.7$ TB DRAM. As the sequencing cost continues to decline much faster than Moore's law \cite{muir2016real}, we envision the need to analyze even bigger metagenomic graphs using distributed-memory algorithms in the near future. In general, it continues to be an important challenge for researchers on how to optimize the distributed-memory parallel algorithms as well as systems to reach the peak performance for irregular problems with high communication to computation ratios.  
% \input{Tables/compareSeq.tex}

Overall, we see that our proposed algorithm and the optimizations help us improve the state-of-the-art for distributed-memory parallel solution to the graph connectivity problem. Simple and fast heuristics to detect the graph structure enables our algorithm to choose the appropriate method dynamically for computing connectivity. This approach enabled us to compute connectivity for a graph with more than 50 billion edges and 300 million components in less than 4 minutes. The speedup we achieve over the state-of-the-art algorithm ranges from 1.1x to 24.5x.  

\section{Conclusion}
In this work, we presented an efficient distributed memory algorithm for parallel connectivity, based on the Shiloach-Vishkin (SV) PRAM algorithm. We proposed an edge-based adaptation of this classic algorithm and optimizations to improve its practical efficiency in distributed systems. Our algorithm is capable of finding connected components in large undirected graphs. We show that a dynamic approach that analyzes the graph and selectively uses the parallel BFS and SV algorithms achieves better performance than a static approach using one or both of these two methods. The dynamic approach prefers BFS execution only for a large short-diameter graph component. 

Our method is efficient as well as generic, as demonstrated by the strong scalability of the algorithm on a variety of graph types. We also observed better performance when compared to a recent state-of-the-art algorithm. The measured speedup is significant, particularly in the case of large diameter graphs.
\section*{Acknowledgment}
We thank George Slota for sharing the implementation of Multistep method and helping us reproduce previous results. We also thank Aydin Bulu\c{c} for making the Combinatorial BLAS library publicly accessible.
This research was supported in part by the National Science Foundation under IIS-1416259.  The cluster used for preliminary experiments in this work was supported by the National Science Foundation under CNS-1229081.

\bibliographystyle{IEEEtran}
\bibliography{References}
\clearpage
\begin{IEEEbiography}
[{\includegraphics[width=1in,height=1.25in,clip,keepaspectratio,frame]{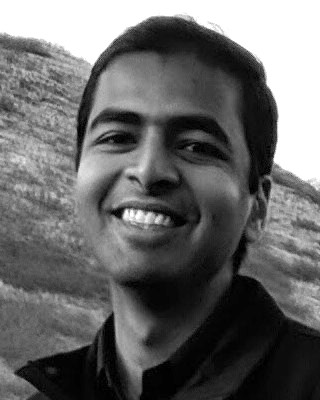}}]{Chirag Jain} received his B. Tech in computer science from Indian Institute of Technology Delhi. He is currently a PhD student in the School of Computational Science and Engineering at Georgia Institute of Technology, Atlanta, USA. His research interests include bioinformatics, combinatorial algorithms and high-performance computing.
\vspace*{10mm}
\end{IEEEbiography}

\vspace{-11ex}

\begin{IEEEbiography}[{\includegraphics[width=1in,height=1.25in,clip,keepaspectratio,frame]{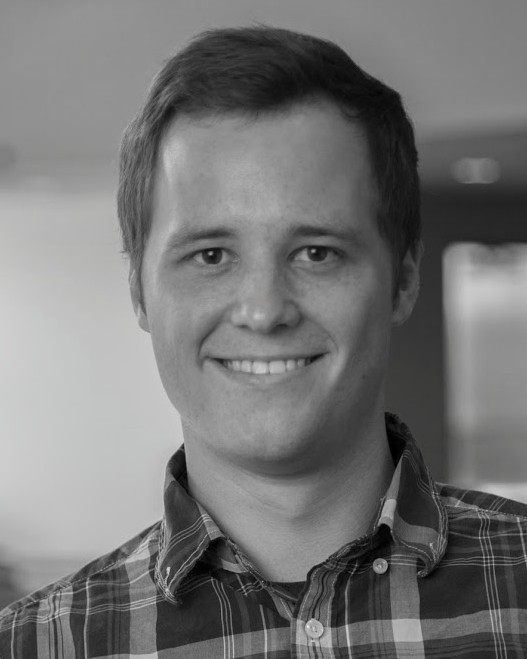}}]{Patrick Flick}
received his Bachelor and Master of Science degrees in computer science from the Karlsruhe Institute of Technology in Germany. Currently, he is a PhD student in Computational Science and Engineering at Georgia Institute of Technology, Atlanta, USA. Patrick's research interests include high performance computing, string algorithms, and graph algorithms.
\end{IEEEbiography}

\vspace{-7.2ex}

\begin{IEEEbiography}[{\includegraphics[angle=270, origin=c,width=1in,height=1.25in,clip,keepaspectratio,frame]{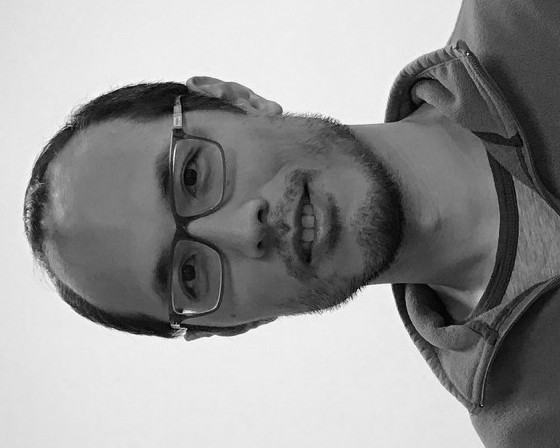}}]{Tony Pan} received Sc. B. in Biophysics from Brown University, and M.S. in Computer Science from Rensselaer Polytechnic Institute.  Previously, he held positions at General Electric, The Ohio State University, and Emory University.  Currently, he is a PhD student in Computational Science and Engineering at Georgia Institute of Technology.  His research interests include high performance computing, distributed information systems, bioinformatics, and biomedical and imaging informatics.
\end{IEEEbiography}

\vspace{-4ex}

\begin{IEEEbiography}
[{\includegraphics[width=1in,height=1.25in,clip,keepaspectratio,frame]{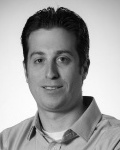}}]{Oded Green}
is a research scientist  in the School of Computational Science and Engineering at Georgia Institute of Technology, where he also received is PhD. Oded received his MSc in electrical engineering and his BSc from in computer engineering, both from the Technion, Israel Institute of Technology. Oded's research focuses on improving performance and increasing scalability for large-scale data analytics using a wide range of high performance computing platforms. 
\end{IEEEbiography}

\begin{IEEEbiography}[{\includegraphics[width=1in,height=1.25in,clip,keepaspectratio,frame]{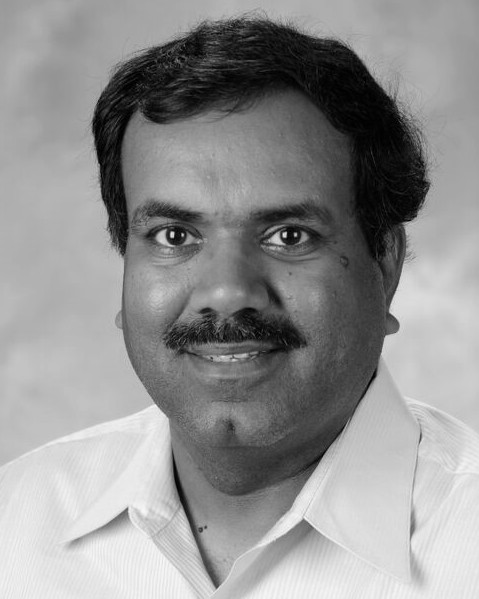}}]{Srinivas Aluru}
is a professor in the School of Computational Science and Engineering at Georgia Institute of Technology. He co-directs the Georgia Tech Interdisciplinary Research Institute in Data Engineering and Science (IDEaS), and co-leads the NSF South Big Data Regional Innovation Hub which serves 16 Southern States in the U.S. and Washington D.C. Earlier, he held faculty positions at Iowa State University, Indian Institute of Technology Bombay, New Mexico State University, and Syracuse University. Aluru conducts research in high performance computing, bioinformatics and systems biology, combinatorial scientific computing, and applied algorithms. He is currently serving as the Chair of the ACM Special Interest Group on Bioinformatics, Computational Biology and Biomedical Informatics (SIGBIO). He is a recipient of the NSF Career award, IBM faculty award, Swarnajayanti Fellowship from the Government of India, and the Outstanding Senior Faculty Research award and the Dean’s award for faculty excellence at Georgia Tech. He received the IEEE Computer Society meritorious service award, and is a Fellow of the AAAS and IEEE.
\end{IEEEbiography}

\end{document}